\documentclass[12pt]{iopart}

\usepackage{epsfig}

\newcommand{\szz}{\sigma_{zz}}

\newcommand{\sxz}{\sigma_{xz}}
\newcommand{\sab}{\sigma_{\alpha\beta}}

\begin{document}

\title{Sensitivity of the stress response function to packing preparation}
\author{
A.P.F.~Atman\dag,
P.~Brunet\dag,
J.~Geng\ddag,
G.~Reydellet\dag,
G.~Combe$^\sharp$,
P.~Claudin\dag,
R.P.~Behringer\ddag\ and
E.~Cl\'ement\dag
}

\address{\dag\ 
Laboratoire de Physique et M\'ecanique des Milieux H\'et\'erog\`enes,
ESPCI, 10 rue Vauquelin, 75231 Paris Cedex 05, France.
}

\address{\ddag\ 
Department of Physics \& Center for Nonlinear and Complex Systems,\\
Duke University, Durham NC, 27708-0305, USA.
}

\address{$^\sharp$ 
Laboratoire Interdisciplinaire de Recherche Impliquant la G\'eologie et
la M\'ecanique, BP 53, 38041 Grenoble Cedex 9, France.
}

\begin{abstract}
A granular assembly composed of a collection of identical grains may pack
under different microscopic configurations with microscopic features
that are sensitive to the preparation history.  A given configuration
may also change in response to external actions such as compression,
shearing etc.  We show using a mechanical response function method
developed experimentally and numerically, that the macroscopic stress
profiles are strongly dependent on these preparation procedures. These
results were obtained for both two and three dimensions.  The method
reveals that, under a given preparation history, the macroscopic
symmetries of the granular material is affected and in most cases
significant departures from isotropy should be observed. This suggests
a new path toward a non-intrusive test of granular material
constitutive properties.

\end{abstract}

\pacs{81.05.Rm, 45.70.Cc, 45.70.-n}

\maketitle

Particular attention has been devoted to the study of the static
properties of granular assemblies over the past 15 years
\cite{reviews}. In that regard, a now famous and paradigmatic example
is that of the stress distribution under a pile of sand. Depending on
the manner this pile is built, different pressure profiles are
measured. For example, when the material is poured from a point
source, e.g. a hopper, then a stress `dip' occurs at the base, under
the apex of the pile \cite{SN81,BHB97,VHCBC99}, whereas when the
material is poured in a rain-like procedure only a slight `hump'
occurs \cite{VHCBC99}. In this context, a discussion on the nature of
the equations governing the distribution of stresses in granular media
has emerged \cite{CWBC98,CWCB98,S98}. The proposition of hyperbolic
equations \cite{BCC95,WCCB96,WCC97} as opposed to the elliptic
equations of elasticity lead to the need for newer more discriminating
tests. These typically involved determinating the stress response to a
localized force for a slab of grains (the response function). The
initial motivation was that the hyperbolic equations predict a double
peaked response (a ring in 3D) which would contrast with a single
peaked profile expected for elliptic equations out of isotropic
elasticity. It is now clear that for generic granular assemblies made
of rough grains, an elastic description using elliptic equations is
most likely correct \cite{RC01,GHLBRVCL01,GG02a}. But it is important
to note that this need not be the case for samples prepared in the
specific case of isostaticity (see \cite{R00,HTW01,BB02,M02,KN04}
and references therein).

One interesting outcome from the previous studies is the fact that, as
the layer geometry is simpler than that of the pile, this stress
response is a powerful tool to explore the effects of preparation
procedures and details of the microstructure. The ultimate goal is to
be able to link the response profiles to the internal texture of the
packing. For instance, the microscopic texture of the packing is quite
sensitive to the details of external actions such as pouring,
shearing, etc... One would like to have a probe of this structure
which determines the effective elastic properties.

In this paper, we focus on the sensitivity of stress response profiles
to the system preparation. First, we
give a summary of recent experimental measurements on two-dimensional
packings made of photoelastic grains \cite{GHLBRVCL01,GRCB03}. Then in
the second section, we show series of experiments on three-dimensional
piling \cite{RC01,SRCCL01}. In the third section we present new data
obtained from numerical simulations of granular layers. All these
stress profiles and the impact on future investigations are finally
discussed in the conclusion.

\section{Two-dimensional photoelastic experiments}

Experiments were carried out on 2D granular systems consisting of disks and
pentagonal particles that were cut from a photoelastic material. In these
experiments, it was possible to measure forces on individual particles
internally -- hence the significant value of the approach. In principle, it is
possible to directly obtain the forces at contacts by solving an inverse
problem for the stress fields internal to the particles that satisfy the
patterns seen in photoelastic images. This is a formidable task to carry out
for large collections of particles. Hence, we chose a simpler approach in
which we calibrated the mean force applied to a particle to the gradient of
the image intensity associated with that particle. Specifically, we determine
the mean force vs. $G^{2}$, where $G^{2}$ is the integral of the square
magnitude of the gradient of the image intensity, integrated over a particle.
The justification for this approach is contained in \cite{GHLBRVCL01,GRCB03}.

\begin{figure}[t]
\begin{center}
\epsfig{file=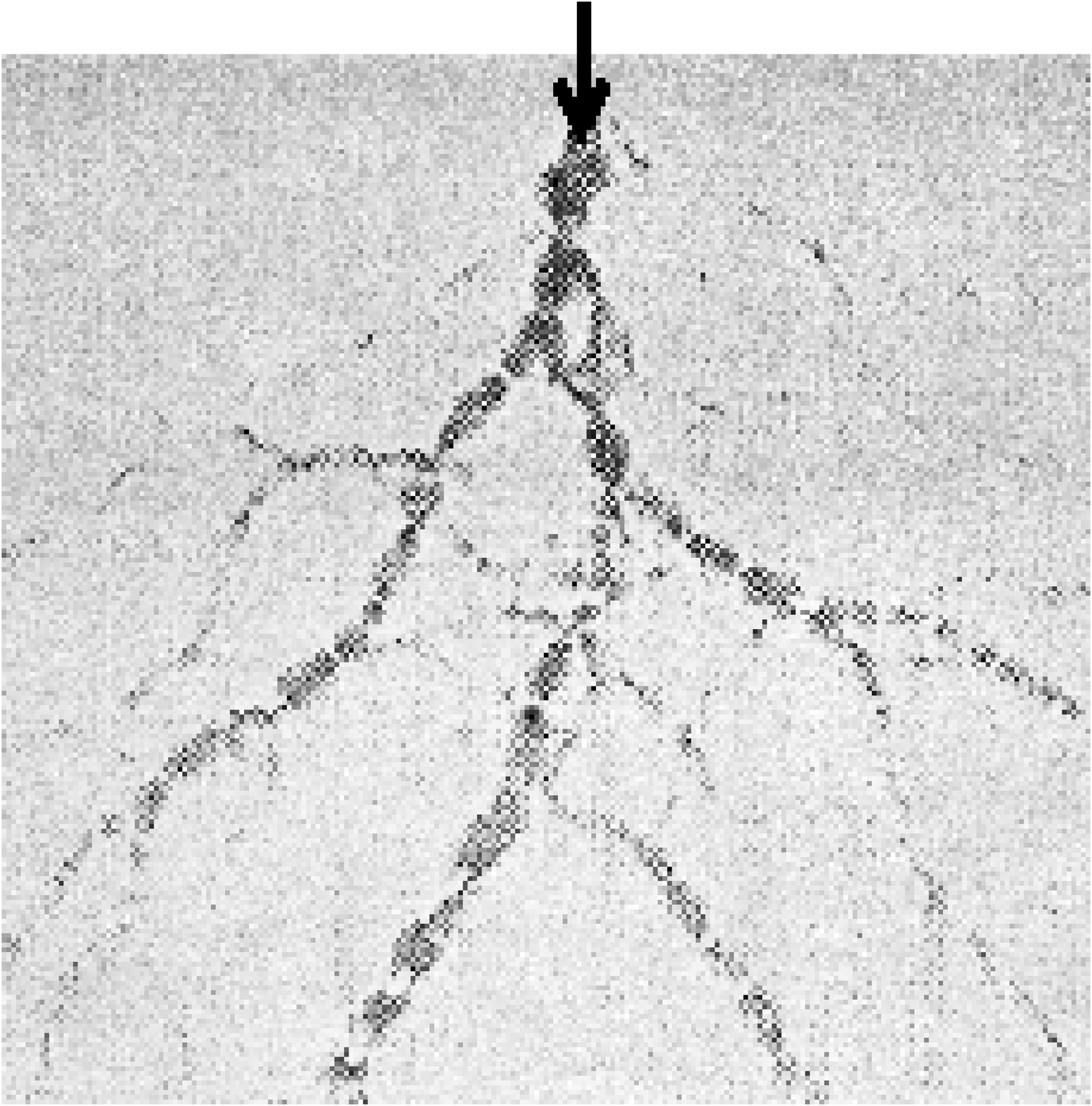,width=0.40\linewidth,angle=0}
\end{center}
\caption{Force distribution in a two-dimensional packing of
photoelastic grains in response to a localized force at the top (the
gravitational part has been subtracted). Darker zones indicate larger
stresses. Chain-like structures are clearly visible. }
\label{fig:forcechains_2D}
\end{figure}

\begin{figure}[t]
\begin{center}
\epsfxsize=0.31\linewidth
\epsfbox{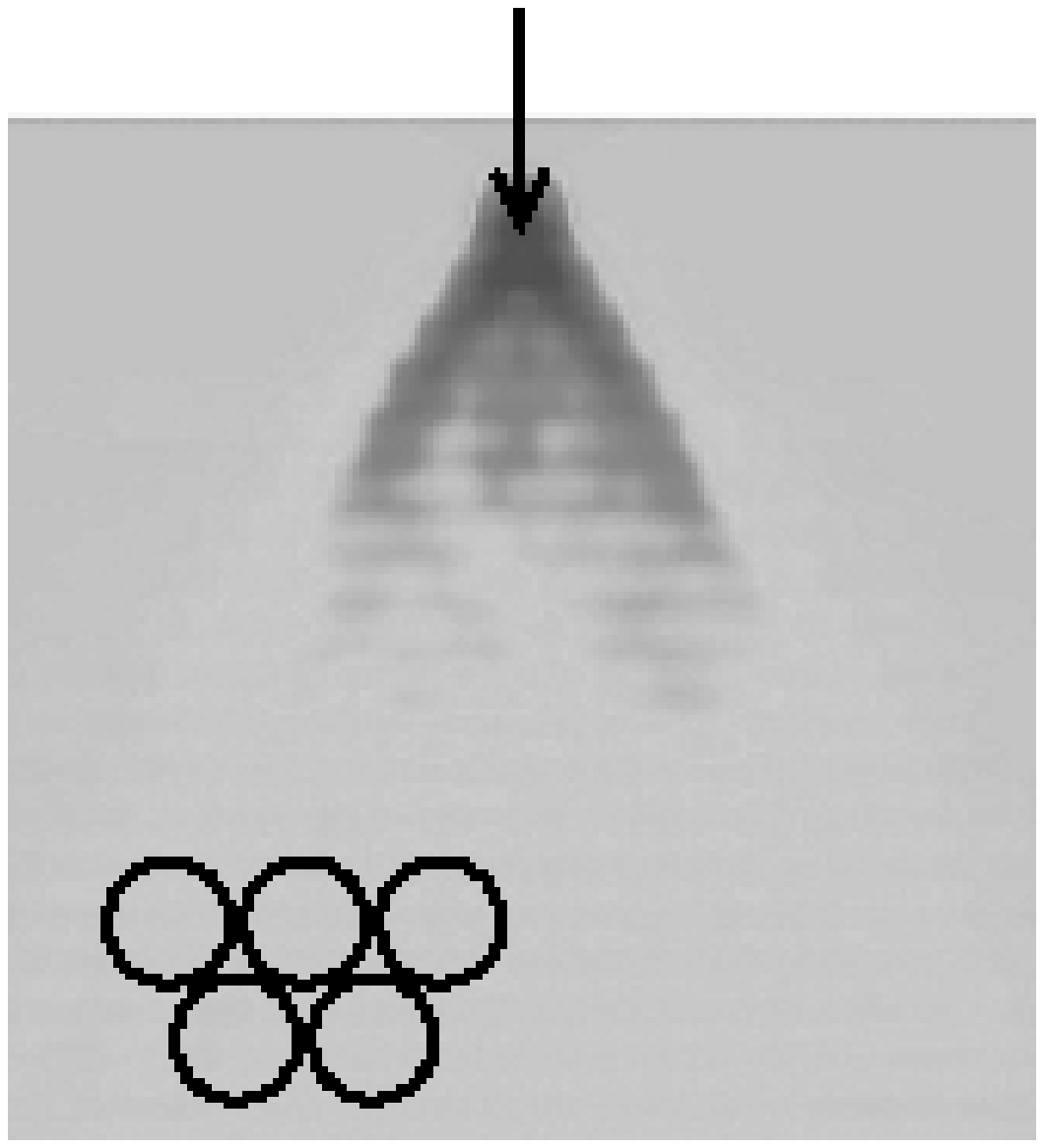}
\hfill
\epsfxsize=0.31\linewidth
\epsfbox{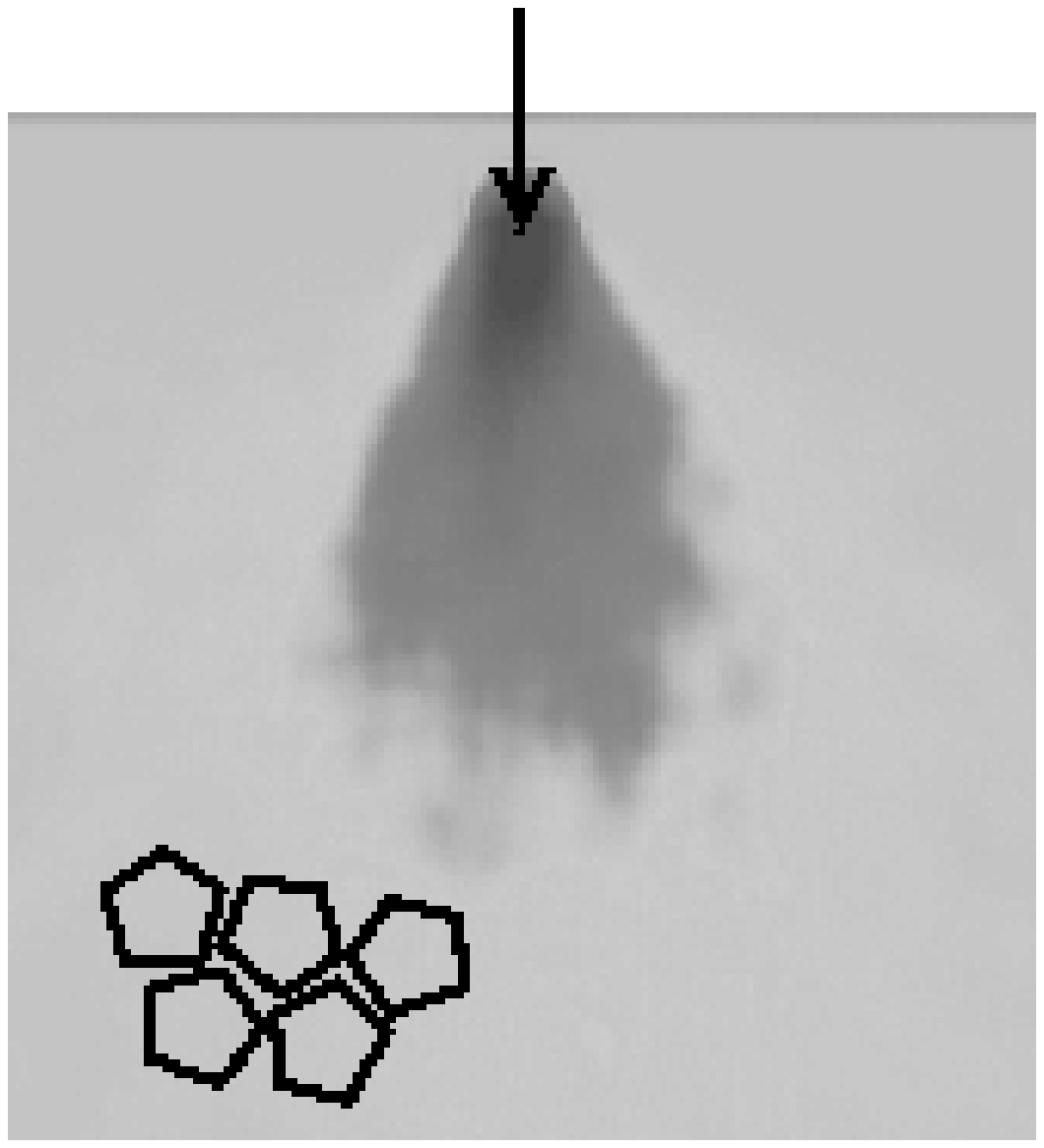}
\hfill
\epsfxsize=0.31\linewidth
\epsfbox{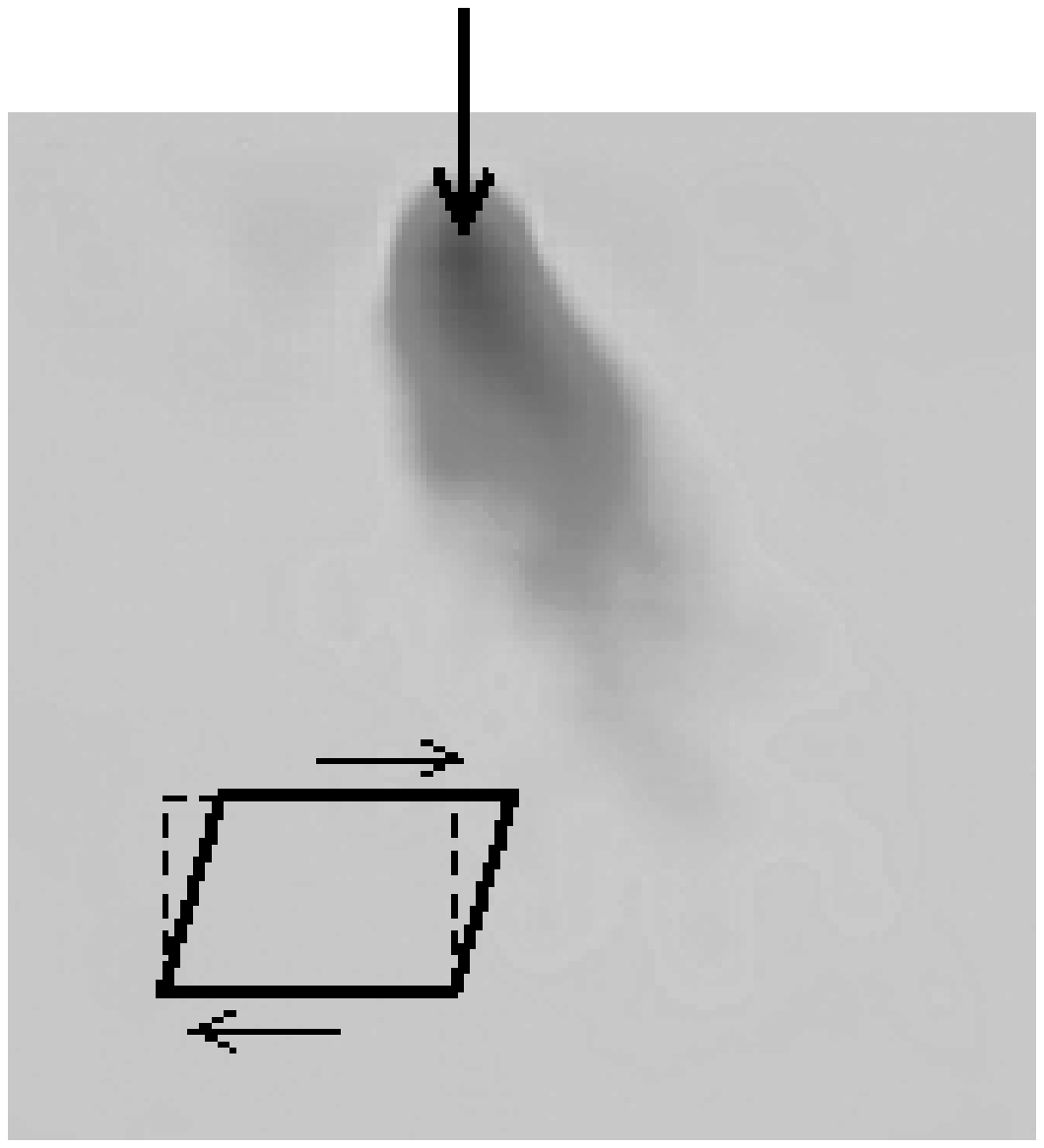}
\end{center}
\caption{Stress response averaged over $\sim50$ trials like that of
figure \ref{fig:forcechains_2D}. The shape of this response shows two
lobes for a regular packing of circular monodisperse beads (left), but
only one lobe (middle) when the layer is disordered (pentagonal
particles). When the layer is sheared beforehand, the response is
skewed in the direction of the shearing (right). The typical height of
these pictures is $15-20$ grain diameters.}
\label{fig:Bob_exp}
\end{figure}

We then carried out a series of experiments with the following general
approach. We created nearly vertical stacks of particles that were
supported very slightly by a vertical glass plate. With this
arrangement, the friction force between particles and the plate was
tiny. Necessarily, this approach caused some gravitational gradients
in the sample, which we removed during the subtraction process
described next. Typical sample sizes contained roughly 1000 particles
and the sample was usually about twice as wide as it was deep.  Each
experiment then consisted in the following: (1) a sample was prepared
as above; (2) images with and without polarizer in place were
obtained; (3) a point force was applied and a third image was
obtained; (4) the point force was removed and a fourth image was
obtained to see if there were shifts in the positions of the
particles. Generally, we rejected an experiment if there were shifts
in the particles positions between the first and fourth images. We
then took the difference between the images with polarizers in places
with and without the applied force.

This then yielded an experimental approximation to the response
function. It is not \emph{a priori} manifest that there is a linear
regime available in these experiments. However, we were able to
identify such a regime in all cases, and we restricted our
observations to this linear regime. An additional point is the fact
that for any given set of macroscopic conditions, such as types of
particles, friction, etc. any given realization was unique, as shown
in figure \ref{fig:forcechains_2D}. Only by obtaining an ensemble of
responses was it possible to generate the mean response
function. Throughout these experiments, we generated ensembles of
typically 50 different realizations for each set of conditions. These
data also yielded information on the variance of the distribution, and
hence the range of local outcomes that one might expect for a given
experiment. In these experiments, we considered a variety of different
conditions. Specifically, we considered ordered hexagonal and square
packing, for which we varied the particle friction coefficient and we
considered completely disordered packing of pentagons. In the last
case, we considered controlled amounts of disorder through bi-disperse
packing of disks. In most of these experiments, we considered weakly
compressed states (compressed due to gravity) and in one set of
experiments, we explored the effect of modest amounts of shear.

Here, we focus on only two issues, namely the role played by spatial
order within the sample, and the effect of modest shear on force
transmission. The effect of order is highlighted by parts (a) and (b)
of figure \ref{fig:Bob_exp}. The left side of this figure shows the
force response in a spatially order hexagonal packing of particles,
and the middle panel shows data for pentagons. In the ordered case,
transmission occurs chiefly along the lattice directions of the
packing, with some friction-dependent widening with depth. This is
consistent with the general predictions of anisotropic elasticity
\cite{GG02a,OBCS03,GG04,GS04} but experimental limitations could not rule
out predictions of hyperbolic models for a weakly disordered system
\cite{CBCW98}. The case of the pentagons shows a single peak that
broadens linearly with depth, consistent with an elastic-like
picture. The linear broadening with depth is not consistent with a
diffusive-like model, such as the $q$-model \cite{LNSCMNW95,CLMNW96}.
The right panel of this figure shows the response to a normal force
applied to a sample that has been subjected to a $5^\circ$ simple
shear deformation.  The shear sets up a strong network of force chains
that align along a direction of $45^\circ$. Indeed, along this
direction, a force-force correlation function suggests long-range
ordering, limited here by the sample size. And, interestingly, the
force response is most strongly refocused to a direction of about
$22^\circ$ with respect to the vertical direction.

In these experiments, we exploited the power of birefringence in
polymeric materials, to directly show the decisive influence of local
granular organization on the macroscopic mechanical properties of a
granular piling. As evidenced by the mean response to a localized
force, this method reveals a connection between local symmetries of
the contact forces and macroscopic anisotropy. On the other hand, the
birefringence method is restricted to 2D.  Thus, it is important to
see whether the effects displayed in 2D would persist for a more
conventional granular packing such as sand. This is the purpose of the
following section.

\section{Three-dimensional experiments}

A priori, the assessment of the stress response to a small localized
force in 3D is a challenging experimental question. First, there is no
easy way to directly probe stresses in the bulk. Second, the direct
estimation of stresses at the boundaries, especially the bottom
boundary, puts us in a situation where the measured stress signal would
always be overwhelmed by the hydrostatic pressure. The method that we
employ here to determine the small force response in the large
hydrostatic background is based on a lock-in detection of the stress
response to a slightly modulated signal added to the main localized
force. This turns out to be extremely sensitive and able to
discriminate the response part of the signal \cite{RC01,SRCCL01}. We
also checked explicitly that the responses obtained are restricted to
a linear regime.

\begin{figure}[b]
\begin{center}
\epsfxsize=0.5\linewidth
\epsfbox{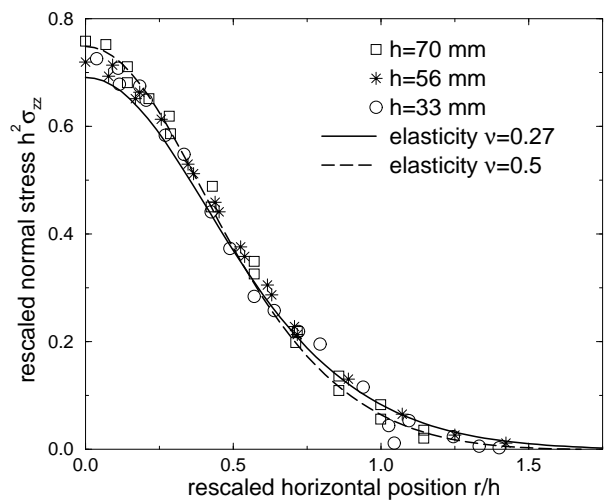} 
\end{center}
\caption{Stress response at the bottom of gelatine layers of various
thicknesses.  The two elastic profiles show the typical accessible
range of shapes for the response of an isotropic layer of finite
thickness on a perfectly rough bottom.  The Poisson coefficient
$\nu=0.27$ gives the widest profile, while $\nu=0.5$
gives the most narrow one \protect\cite{SRCCL01}.}
\label{fig:gelatine}
\end{figure}

An interesting test of our method is provided by the study of a model
elastic material such as gelatine with a known Poisson ratio
$\nu=0.5$. It is first important to note that the necessary presence
of boundaries for all practical purpose is an element that needs to be
considered explicitly in an elastic analysis. In the case of the
stress profiles of an isotropic semi-infinite slab, the response
function is solely dependent on the spatial dimensions but this nice
property is lost for finite depth elastic layers. For example,
boundary roughness and Poisson ratio appear directly in the isotropic
elasticity solution \cite{SRCCL01}. The gelatine results are plotted
on figure \ref{fig:gelatine}. The data have been rescaled by the
thickness, $h$, of the layer and by the amplitude of the extra
force. The comparison to isotropic elasticity is excellent. Note,
however, that the sensitivity of the response to the exact value of
the Poisson coefficient is not particularly high; for instance,
solutions with $\nu=0.27$ and $\nu=0.5$ show very similar shapes.

\begin{figure}[t]
\begin{center}
\epsfxsize=0.5\linewidth
\epsfbox{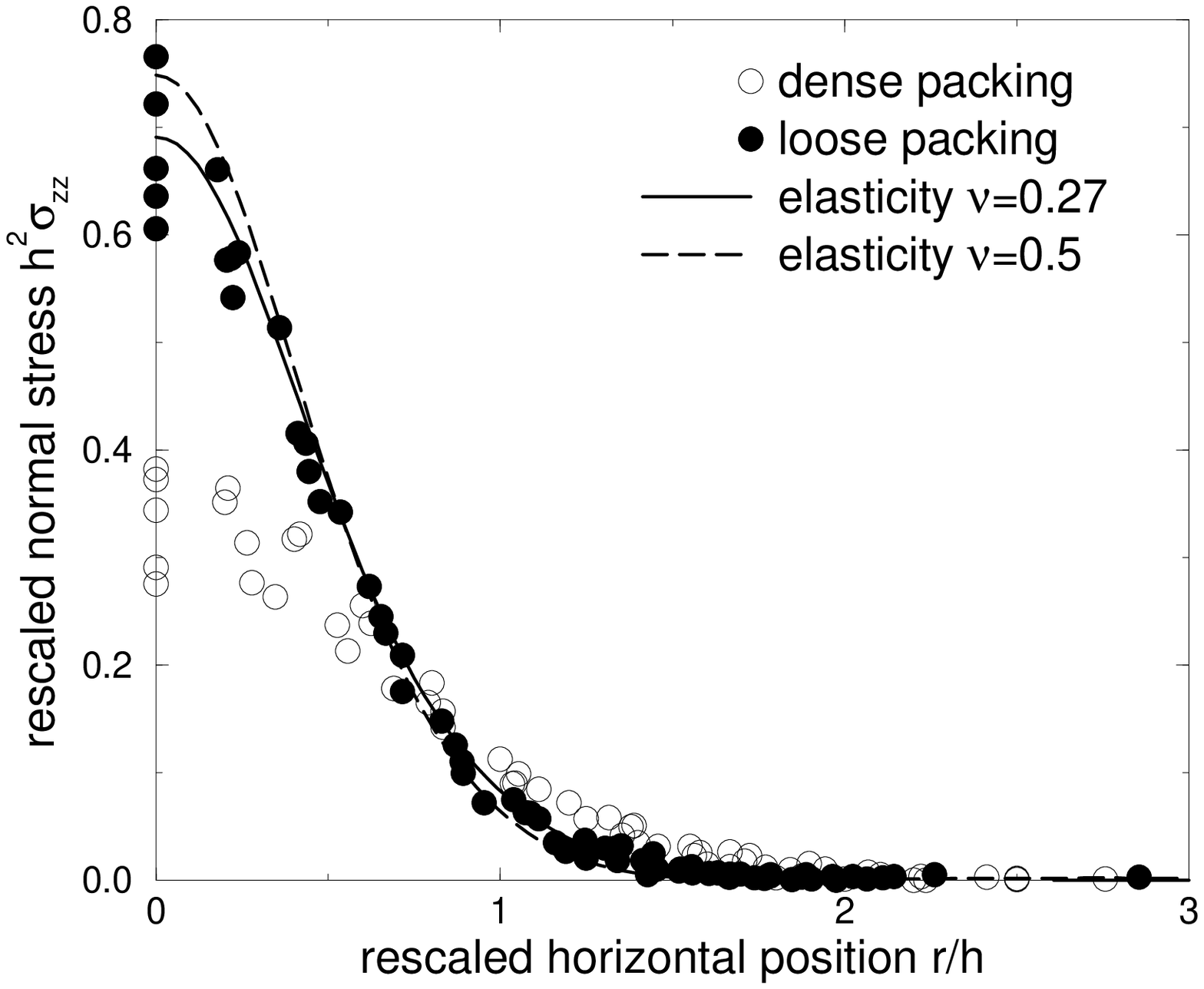}
\hfill
\epsfxsize=0.35\linewidth
\epsfbox{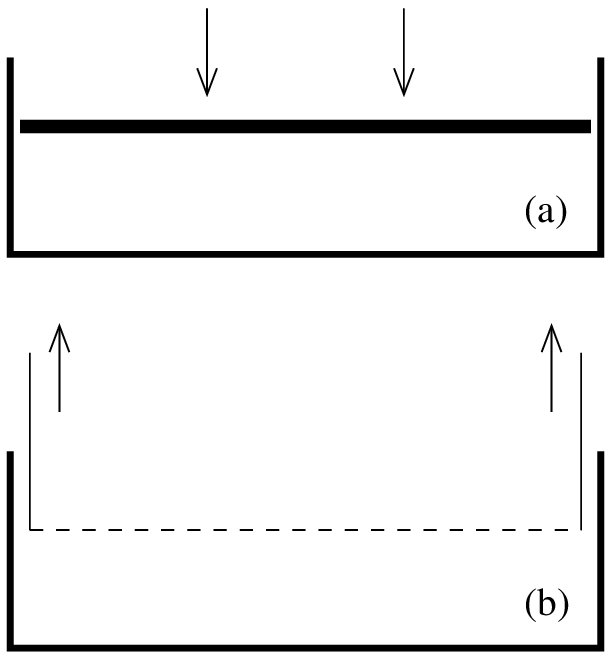}
\end{center}
\caption{Response profiles for the vertical stress, $\szz$. The open
circle data points were obtained on dense, compressed layers of grains
(a), while the filled ones are from rather loose packings (b). We have
plotted together data for layers whose thickness varies from $h \sim
30$ to $h \sim 60$ mm.  We observe that the response function of a
loose piling is much narrower than that of a compressed one. As for
figure \protect\ref{fig:gelatine}, the range of elastic predictions is
shown by the two profiles $\nu=0.27$ and $\nu=0.5$.}
\label{fig:GR_exp}
\end{figure}
\begin{figure}[t]
\begin{center}
\epsfxsize=0.5\linewidth
\epsfbox{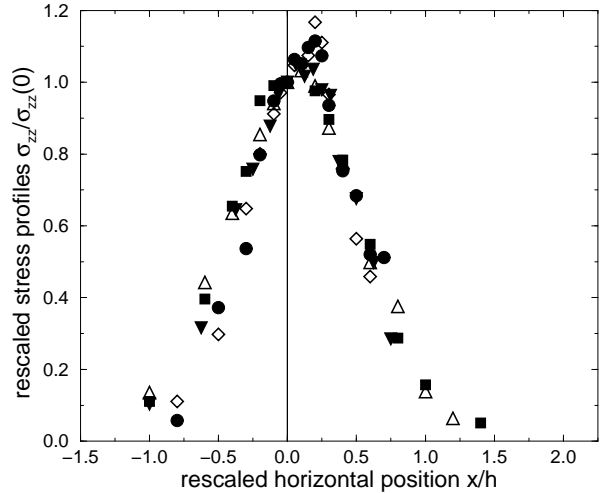}
\end{center}
\caption{Response profile for the vertical stress $\szz$ at the bottom
of a sand layer which has been uniformly sheared along the $x$
axis. The different symbols are for layers of various thicknesses
($50$ to $100$ mm) and shear strains ($2{{}^\circ}$ and
$5{{}^\circ})$. The maximum of the profile is clearly shifted in the
direction of shear. Isotropic elasticity is not able to reproduce such
an asymmetrical shape of the profile with a vertical overload.}
\label{fig:PB_exp}
\end{figure}

We now turn to response functions obtained with different preparations
of Fontainebleau sand ($d \sim 300 \mu$m). To date, we have tested
granular assemblies using the four following preparation methods: (i)
a loose packing, (ii) a dense packing, both cases having the direction
of gravity as a symmetry axis \cite{SRCCL01}, (iii) a sheared granular
assembly and, (iv) an horizontal slab prepared using successive
avalanches \cite{ABRCCB04}. We present in figures \ref{fig:GR_exp} and
\ref{fig:PB_exp} cases (i), (ii) and (iii). In all situations a linear
elastic-like rescaling of the response with depth $h$ was obtained --
the scaling here refers to the fact that elastic stress fields have
the form $\sab = 1/z^p f(x/z)$ ($p=2$ in 3D, $p=1$ in 2D).

Except for case (i), the predictions of isotropic elasticity are more
or less inconsistent with our experimental results. For instance, the
dense packing prepared by successive compression of deposited layers
gives a response which is far too wide to be well fitted by such a
model. For preparations obtained either in a shearing box or by
avalanches -- cases (iii) and (iv) -- a shift of the response maximum
with respect to the vertical direction below the applied force is
present. It directly attests to the presence of an anisotropic medium
with a symmetry axis other than the vertical direction. From all these
data, as in 2D, we conclude the existence of a macroscopic anisotropy
induced by the pile preparation. Similar conclusion could be drawn for
3D response functions experiments \cite{MJN02,SMJN03} or simulations
and calculations performed on highly symmetrical crystalline packing
\cite{GG02a,BCCZ02,OP04,SVG02,GG04b}.

Recently, there have been attempts to describe granular assemblies
using orthotropic elasticity \cite{OBCS03,ABRCCB04}, where the
direction of larger compression is
taken as the stiff direction and thus the main symmetry axis of the
model. This approach can provide a good qualitative account of all the
experimental results but the number of parameters involved is too high
to provide a decisive interpretation.  Therefore, there is a crucial
need to measure other quantities based on the response function
method, such as shear stresses or responses to a non-vertical force as
in \cite{GRCB03}. This is also the object of a numerical study in the
following section. A large collection of these experimental data can
provide severe tests of any proposed mechanical modelling of granular
assemblies under quasistatic deformation. We pursue this issue further
via a numerical study which we discuss in the following section.

\section{Numerical simulations}

In order to complement the experimental data presented above, we
performed extensive 2D simulations of various grain packings obtained
with different preparation methods. The control of all the parameters
of the simulations as well as the ability to measure both micro (grain
size) and macro (system size) quantities ensures a useful feedback to
the experiments. In addition, a numerical approach allows an easier
exploration of a larger range of preparation procedures.  Similar
numerical studies have been recently carried out, either on granular
packings \cite{M00,GWM04} or on amorphous elastic bodies such as
Lennard-Jones particles \cite{LTWB04}.

\subsection{Simulation model and parameters}

The system consists of a set of $N$ polydisperse discs, with random
radii homogeneously distributed between $R_{min}$ and
$R_{max}=2R_{min}$.  This system is simulated using a discrete element
method -- molecular dynamics (MD) with a third order
predictor-corrector scheme \cite{AT87}. 
The rheology of contacts is that of Kelvin-Voigt.
All grains interact via a linear elastic law and Coulomb friction when
they are in contact. The normal contact force $f_n$ is related to the
normal apparent interpenetration $\delta$ of the contact as $f_n = k_n
\cdot \delta$, where $k_n$ is a normal stiffness coefficient
($\delta > 0$ if a contact is present, $\delta = 0$ if there is no
contact). Note that this linear behaviour is consistent with Hertz law for
discs. The tangential component $f_t$ of the contact force is
proportional to the tangential elastic relative displacement, with a
tangential stiffness coefficient $k_t$. The Coulomb condition $\vert
f_t \vert \le \mu f_n$ requires an incremental evaluation of $f_t$
every time step, which leads to some amount of slip each time one of
the equalities $f_t = \pm \mu f_n$ is imposed. A normal viscous
component opposing the relative normal motion of any pair of grains in
contact is also added to the elastic force $f_n$ to obtain a critical
damping of the dynamics.  We chose a value $\mu = 0.5$ for both the static
and dynamic friction coefficients between the grains. All grains can
be considered as quite rigid, $\kappa = k_n / P = 1000$, where $\kappa$
is a dimensionless parameter which expresses the level of contact
deformation and where $P$ is a pressure related to the weight of
grains assembly. The ratio $k_t / k_n$ was set to $0.5$ or $0.7$
\cite{SDW96}. The bottom of the system is a horizontal line of
similar but fixed grains, with intergranular distances chosen in order
to avoid grains escaping from the system. We use horizontally periodic
boundary conditions.  Finally, we have carried out simulations on
eight different samples of $N=3600$ particles.

\begin{figure}[t]
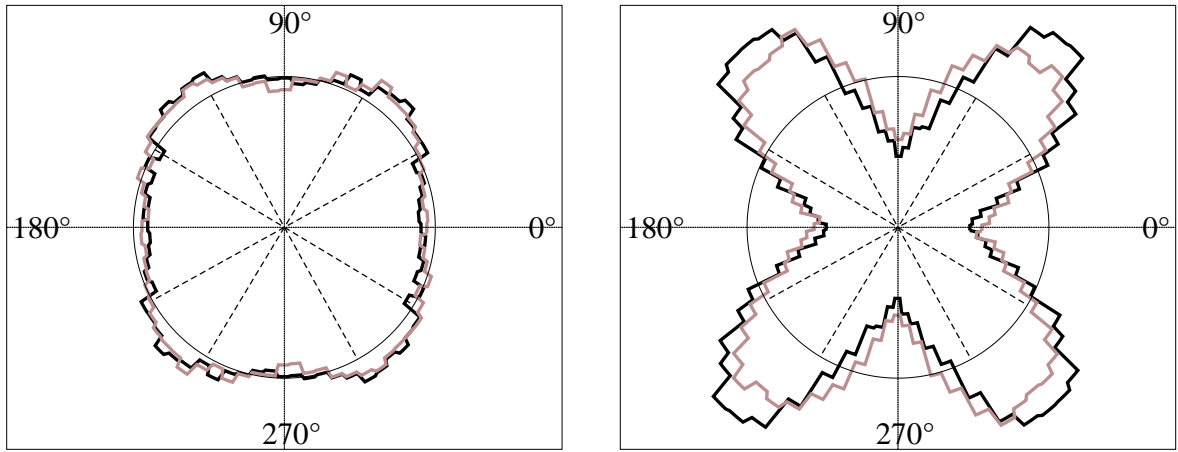

\begin{center}
\epsfig{file=Pdetheta.rain.eps,width=0.48\linewidth,angle=0}
\hfill
\epsfig{file=Pdetheta.GbyG.eps,width=0.48\linewidth,angle=0}
\end{center}
\caption{Contact and force angle distribution for the rain-like (left)
and grain by grain deposition (right) procedures. Black lines are for
the angles of contact whereas gray ones are for the orientation of the
forces -- all angles are measures with respect to the horizontal
axis. The circles are guidelines indicating an isotropic
distribution. The texture anisotropy of the grain-by-grain deposition
is striking. These statistics have been obtained over about 50000 (RL)
or 32000 (GG) contact pairs, and the bin width for each histogram
corresponds to an angular interval of $5^\circ$}
\label{fig:Pdetheta}
\end{figure}

We considered two different system preparations. The first is a
rain-like (RL) procedure where all grains are initially placed at the
nodes of a $1$ to $4$ aspect ratio grid. When the deposition starts,
all the constraints are removed at the same time, letting the
particles settle down under gravity with random initial velocities. In
the second preparation scheme, each grain is deposited individually
(grain by grain, or GG) with no initial velocity, by choosing a random
initial position in contact with a grain already present at the outer
surface. The next particle is deposited after 100 MD time steps if all grains
already present have at least one contact. The equilibrium criteria consists
of the five following tests which are applied after each period of $100$ MD
time steps: (1) the number of gained/lost contacts between particles
during this period is zero, (2) similarly, the number of sliding
contacts is also zero, (3) the integrated force measured at the bottom
of the layer is equal to the sum of the weight of all the grains
within a minimal tolerance, (4) all the particles have formed at least
two contacts, and (5) the total kinetic energy is lower than some low
threshold.

\begin{figure}[t]
\begin{center}
\epsfig{file=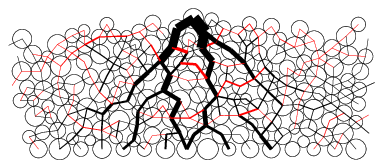,width=0.70\linewidth,angle=0}
\end{center}
\caption{Contact forces in response to an incremental small localized
force at the top surface.  Lines are bolder for larger forces. Black
(red/grey) lines indicate contacts where the force has increased
(decreased). The numerical results presented in this paper have been
obtained with assemblies of $3600$ particles forming typical layers of
aspect ratio $1 \times 4$, of 21 grains thickness, and with horizontal
periodic boundary conditions.}
\label{fig:forcechains_numerics}
\end{figure}

We then carried out a microscopic characterization of the deposited
layers and computed such quantities as the contact and force angle
distributions.  These are plotted in polar representation in figure
\ref{fig:Pdetheta}. It is striking to see the difference between the
two plots: the first one is close to a circle (quasi-isotropic
distribution) whereas the second one shows four distinct
lobes. Similar features have been observed experimentally by Calvetti
\emph{et al.} \cite{CCL97} on a Schneebeli assembly \cite{S56}
prepared using something like the GG procedure, and also by Radjai
\emph{et al.}  \cite{RWJM98,RRM99,RTR03} with discrete element method
simulations (Contact Dynamics). In the same vein,
Geng \emph{et al.} \cite{GLBH01} found
that a raining procedure for building 2D heaps of photoelastic
particles resulted in more nearly axisymmetric distributions of
contacts than did a point-source pouring method.  Interestingly, in
these latter experiments, the underlying symmetry was hexagonal rather
than square.  At the system size level, the difference is less impressive:
we have computed the solid fraction $\phi$ and found
$\phi_{RL}=0.820(2)$ and $\phi_{GG}=0.812(4)$ respectively. The
average numbers of contacts per grain, $Z$, are also rather close:
$Z_{RL}=3.39(1)$ and $Z_{GG}=3.517(8)$.

\subsection{The response function}

After the deposition procedure converges to a static equilibrium,
which typically takes $10^6$ MD time steps, an additional force
$F_{0}$ localized on a single grain is added at the top of the
layer. Its value is typically of the order of a few times the weight of
a grain. As we emphasize below, $F_{0}$ can be either vertical or
inclined at a $45^\circ$. This force is applied progressively over
$10^{4}$ MD time steps, and we checked that this overloading does not
lead to important rearrangements of the packing. More precisely, no
contacts are either broken or created, and the few observed sliding
events do not push the response out of a linear and reversible
regime. Figure \ref{fig:forcechains_numerics} shows a map of the
contact forces in response to this extra force -- the force network
before the overloading was subtracted. In order to get the averaged
stress response function, several tens of such independent overloads
were performed on assemblies of $3600$ particles. The stress
components $\szz$ and $\sxz$ are measured at the bottom of the system
in the following manner. Like the experimental pressure probe of size
$l$, the stress component $\sigma_{\alpha z}$ at position $x$ is
obtained as the sum over the $\alpha$-components of the forces exerted
on the bottom grains located at horizontal distances between $x-\ell/2$
and $x+\ell/2$ from the overload point. This sum is divided by the
coarse graining length, $\ell$, and by the number of samples $\mathcal{N}$. It is
also normalized by the force $F_0$. Finally:
\begin{equation}
\sigma_{\alpha z} (x) = \frac{1}{\mathcal{N} F_0 \ell} \,
                       \sum_{x-\ell/2<\xi<x+\ell/2} F_\alpha (\xi).
\end{equation}
We checked that this procedure
yields results consistent with a more general definition of the stress
tensor $\sab$ \cite{GG02a,GG04,CMN81,RS81,GG02b}. We note that the
stress components become independent of the coarse graining length
only when $\ell$ is large enough compared to the mean grain diameter
$d$ \cite{GG02b}. Here we used $\ell \sim 7.5d$, which is about the
crossover value.

\begin{figure}[t]
\begin{center}
\epsfig{file={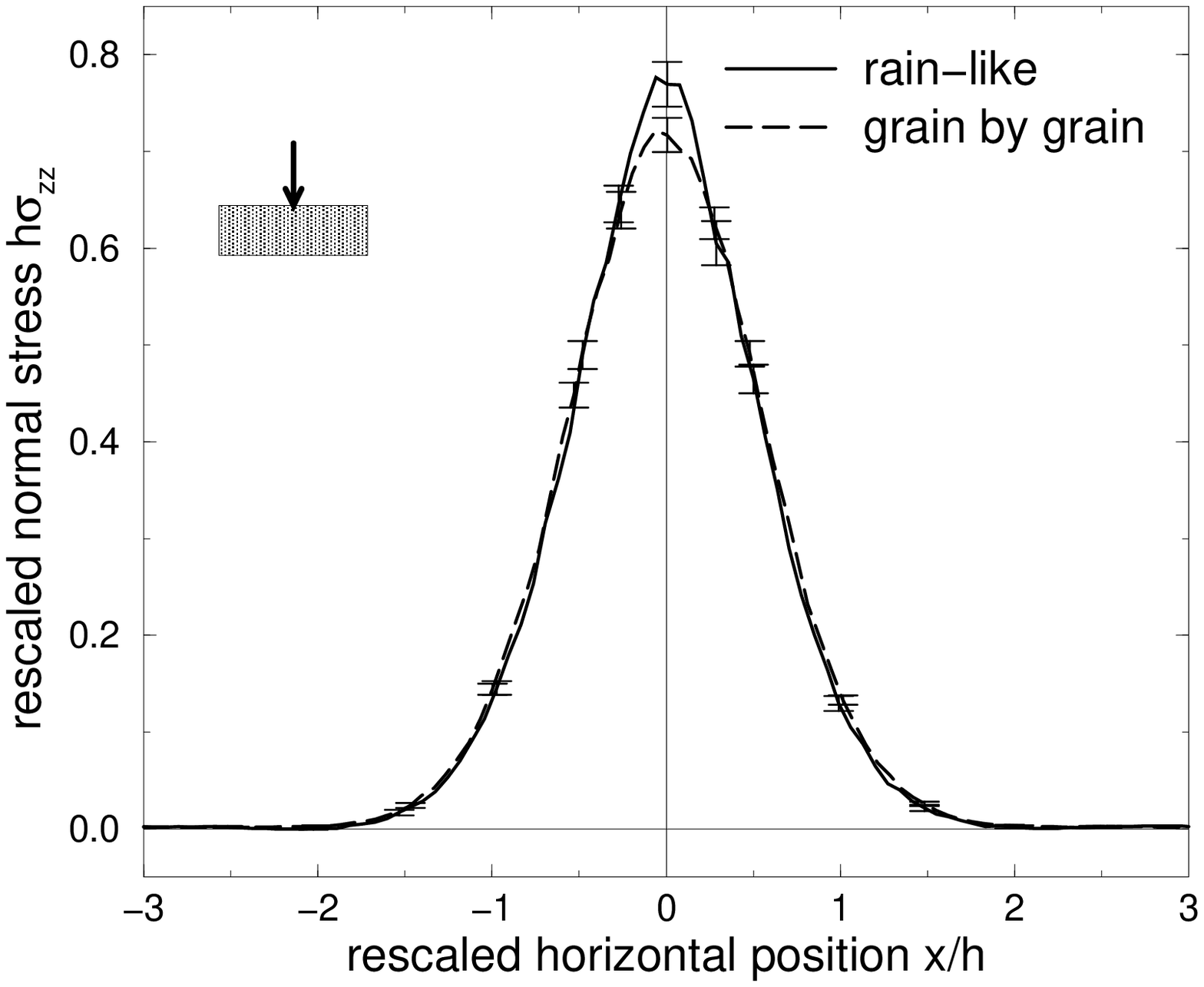},width=0.474\linewidth,angle=0}
\hfill
\epsfig{file={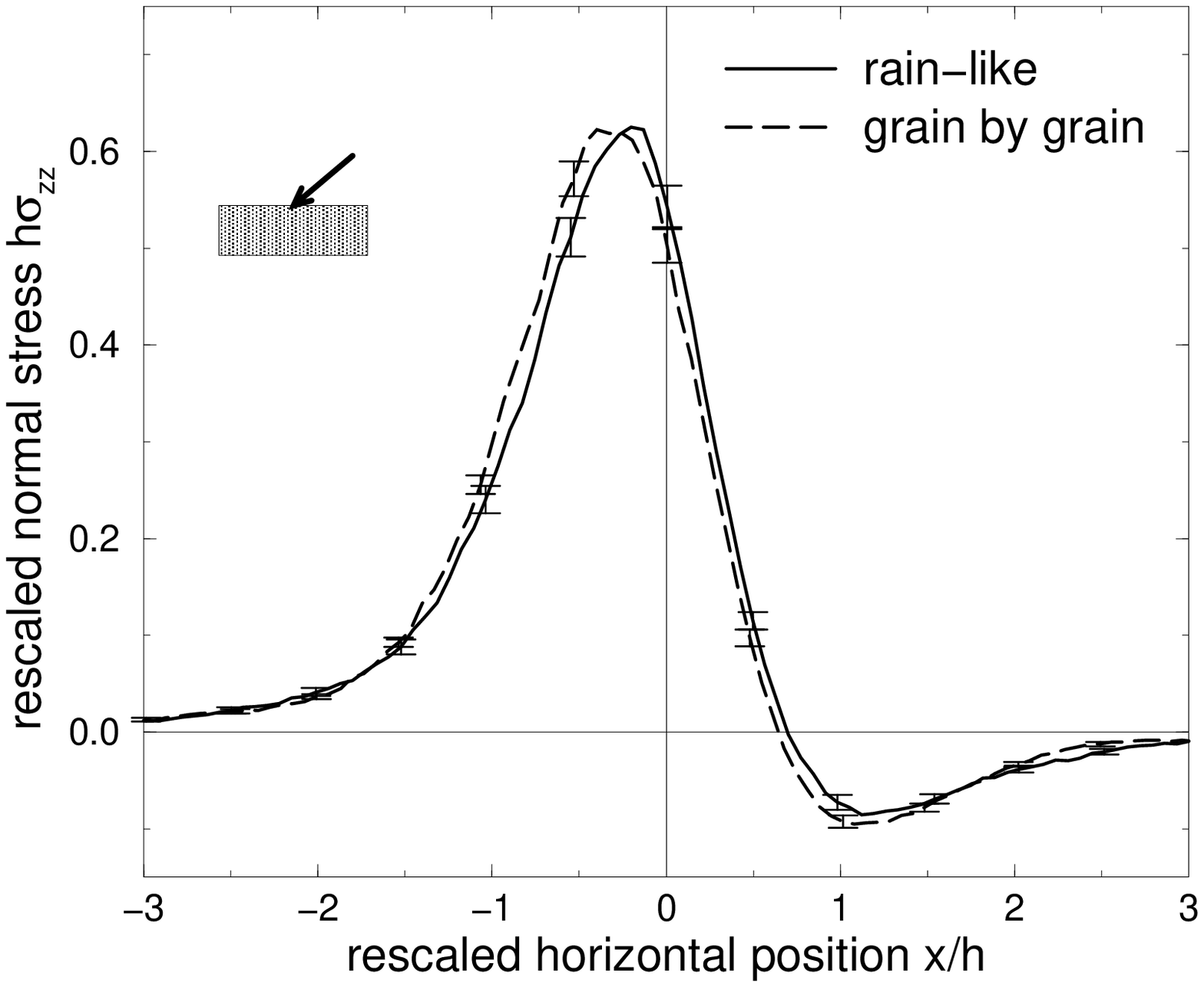},width=0.486\linewidth,angle=0}
\end{center}
\caption{Profiles for the vertical normal stress, $\sigma_{zz}$, in
response to a vertical (left) or a $45^{o}$ inclined (right) top
force. The two preparation procedures show distinct but qualitatively
similar profiles.}
\label{fig:szz_numerics}
\end{figure}
\begin{figure}[t]
\begin{center}
\epsfig{file={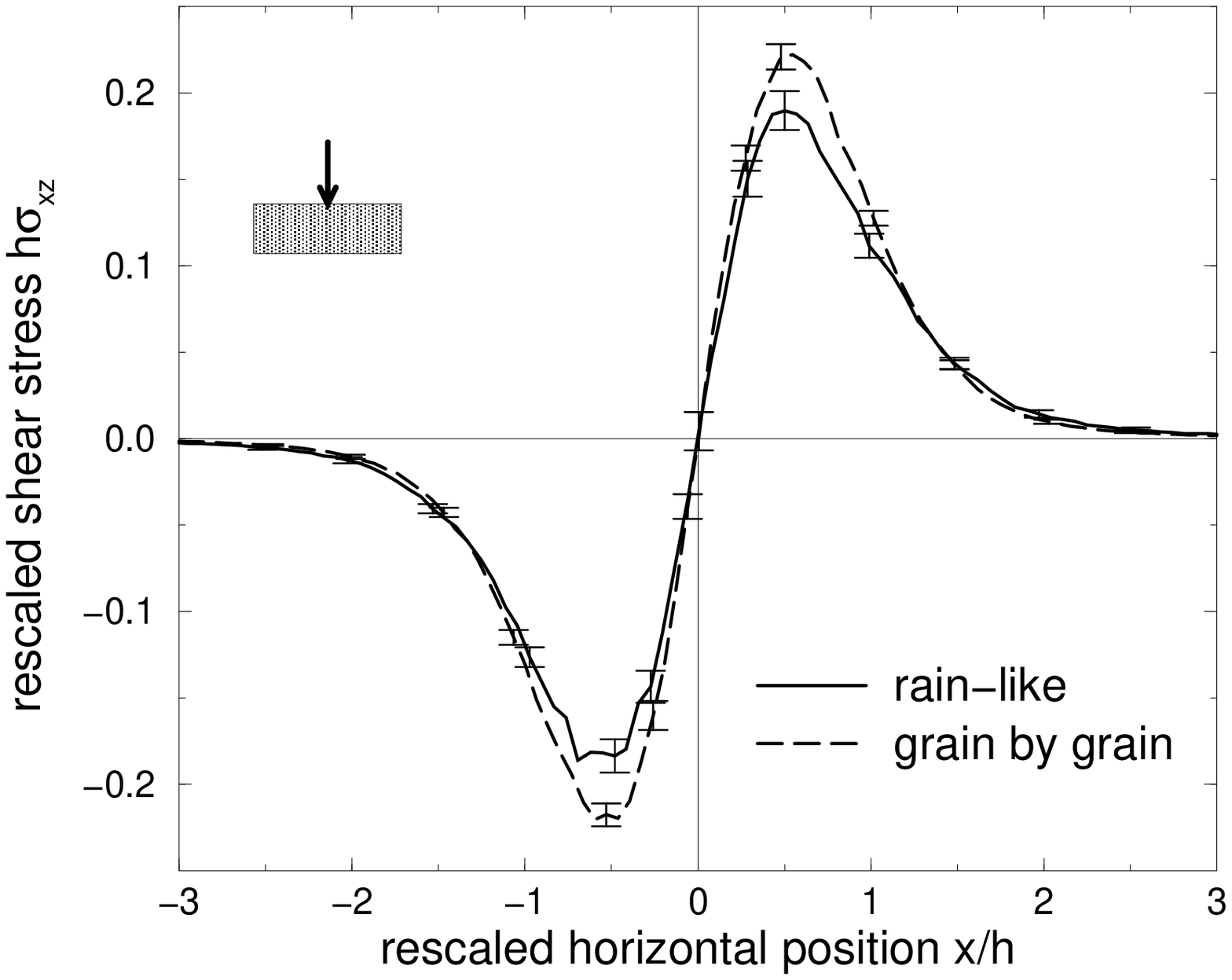},width=0.48\linewidth,angle=0}
\hfill
\epsfig{file={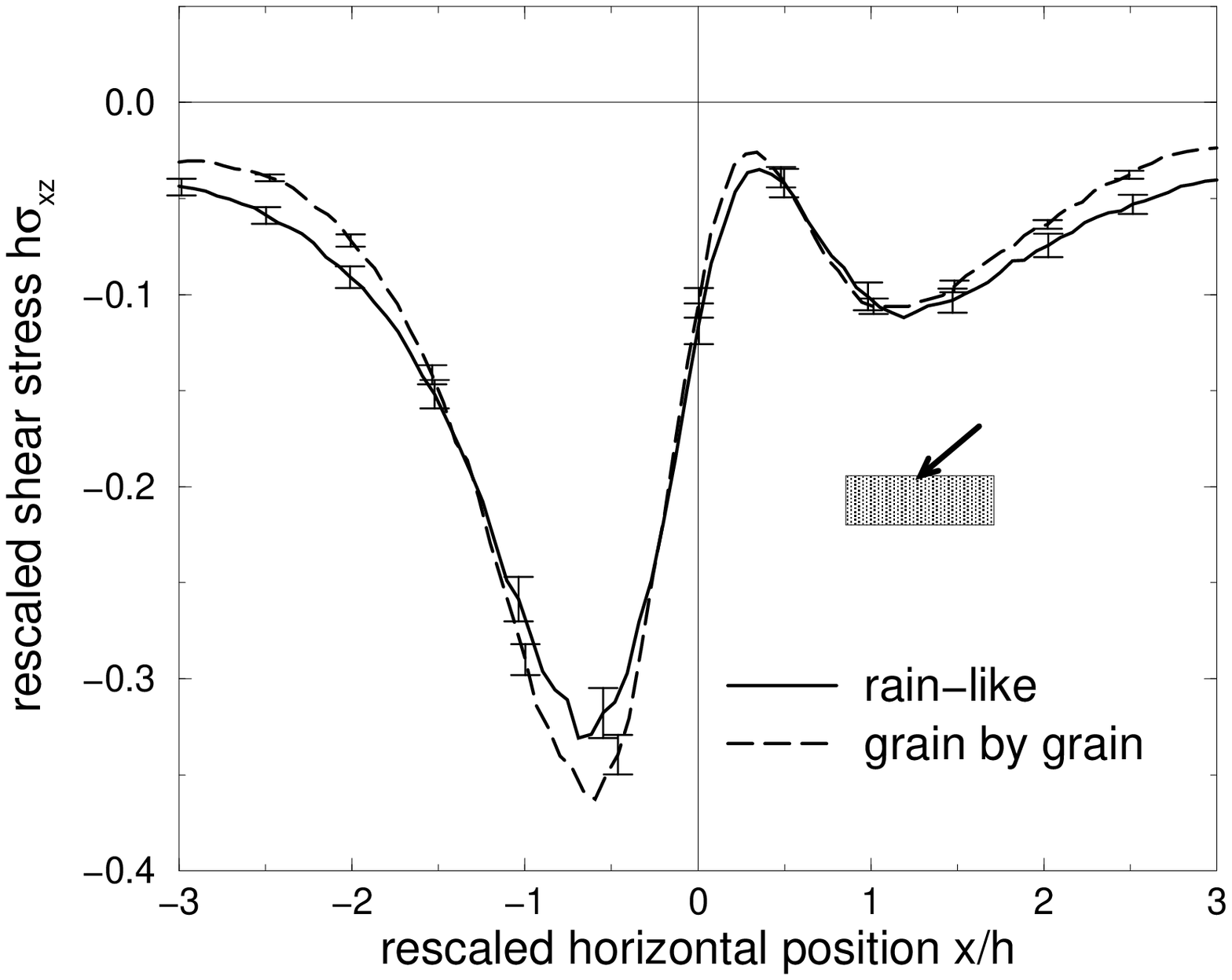},width=0.48\linewidth,angle=0}
\end{center}
\caption{Shear stress $\sigma_{xz}$ profiles in response to a vertical (left)
or a $45^{o}$ inclined (right) top force. The difference between the two
preparation procedures is much more visible than in the data for $\sigma_{zz}$.
In particular, the amplitude of the shear maximum is $\sim 10\%$ larger
for the more anisotropic case. }
\label{fig:sxz_numerics}
\end{figure}

Figures \ref{fig:szz_numerics} and \ref{fig:sxz_numerics} show the
different stress response profiles. They exhibit a single broad
peak.  Although not shown here, and in accordance with the
experimental data presented in the previous sections, these profiles
also scale linearly with the layer thickness. The $\szz$ profiles
measured from the RL or the GG procedures are distinct but
close. Interestingly, the difference between the two preparations is
much more visible on the $\sigma_{xz}$ plots. In particular, the
amplitude of the shear maximum is typically $10\%$ larger in the
more anisotropic case (GG).

\subsection{Comparison with isotropic elasticity}

As in the loose and dense cases of the previous section, both
rain-like and grain-by-grain preparations have the vertical axis as an
axis of symmetry. It is not clear {\em a priori} that any piling
obtained under such conditions would ever be isotropic. From a
microscopic point of view, an inspection of the contact distribution
(figure~\ref{fig:Pdetheta}) suggests that that the RL procedure is
more likely to yield an isotropic effective medium than the GG
deposition procedure. The question is how much are the texture
properties reflected in macroscopic features. Here, we simply try to
adjust the stress response functions predicted by a isotropic
elasticity, and we seek to identify which features of these functions
would be most sensitive for identifying a departure from isotropy.

\begin{figure}[t]
\begin{center}
\epsfig{file={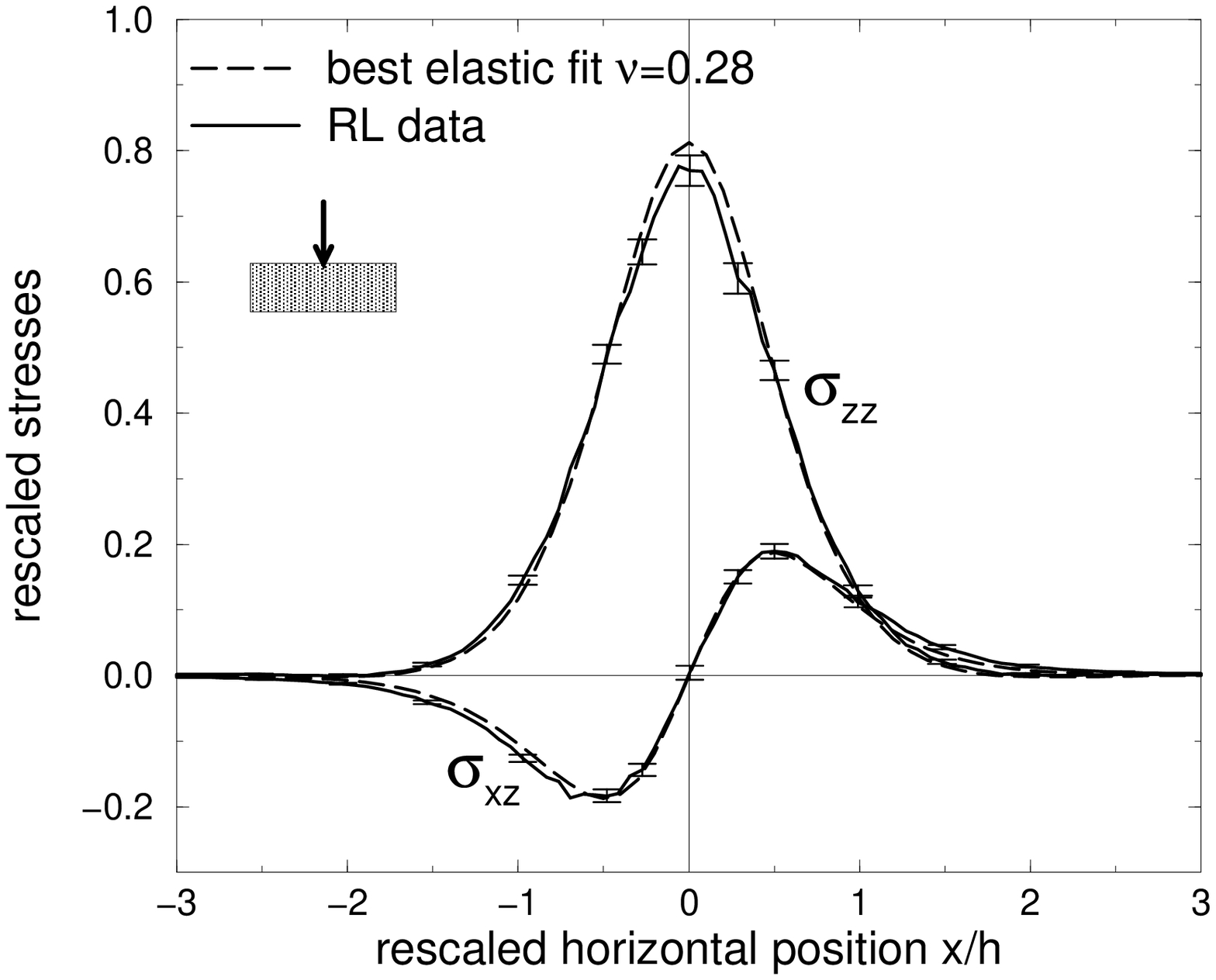},width=0.48\linewidth,angle=0}
\hfill
\epsfig{file={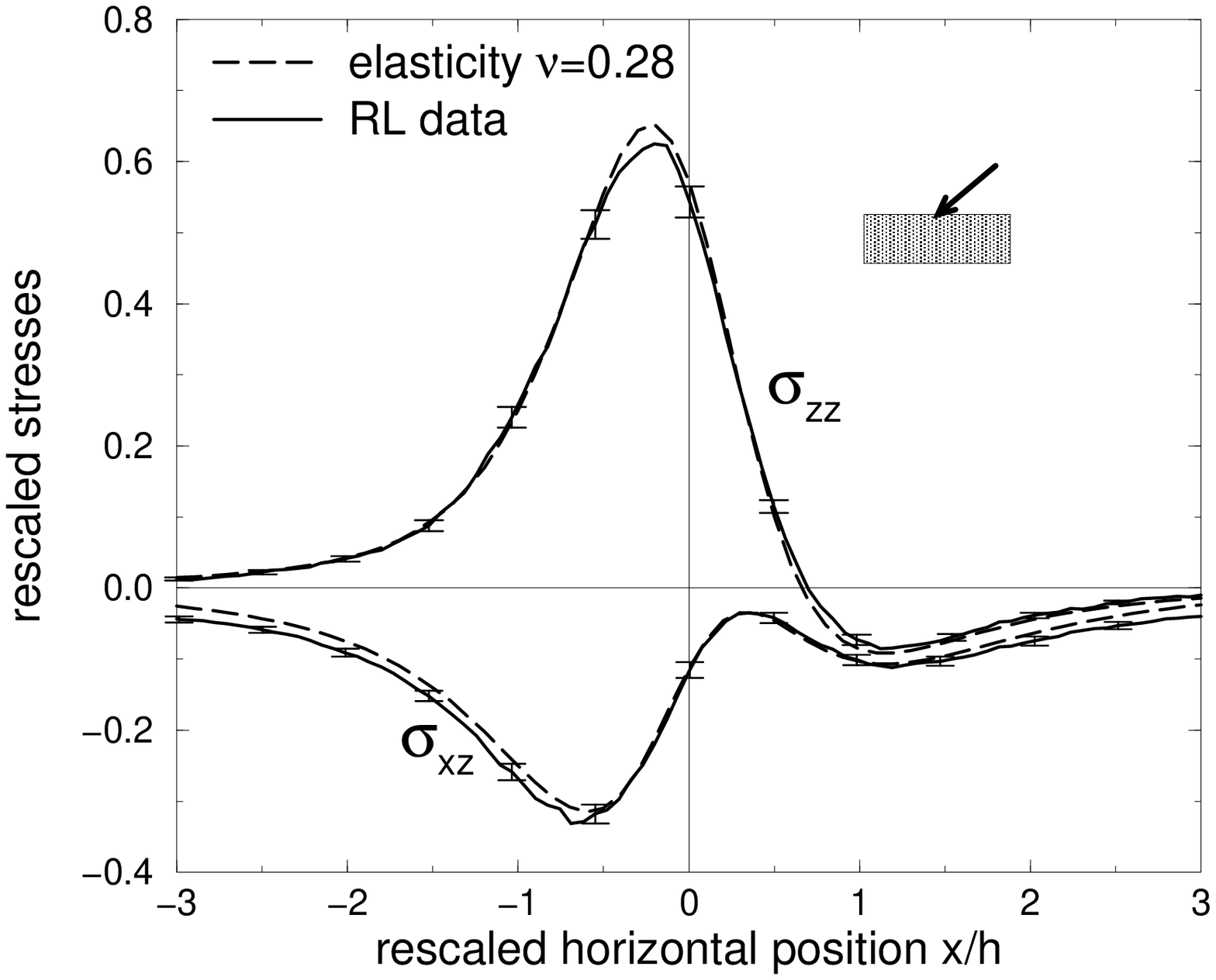},width=0.48\linewidth,angle=0}
\end{center}
\caption{Comparison of the stress profiles with isotropic elasticity.
The fit for both cases of a vertical and an inclined overload
$F_0$ is quite good. These plots are for the rain-like preparation
of the granular layer.}
\label{fig:fitsijRL}
\end{figure}
\begin{figure}[t]
\begin{center}
\epsfig{file={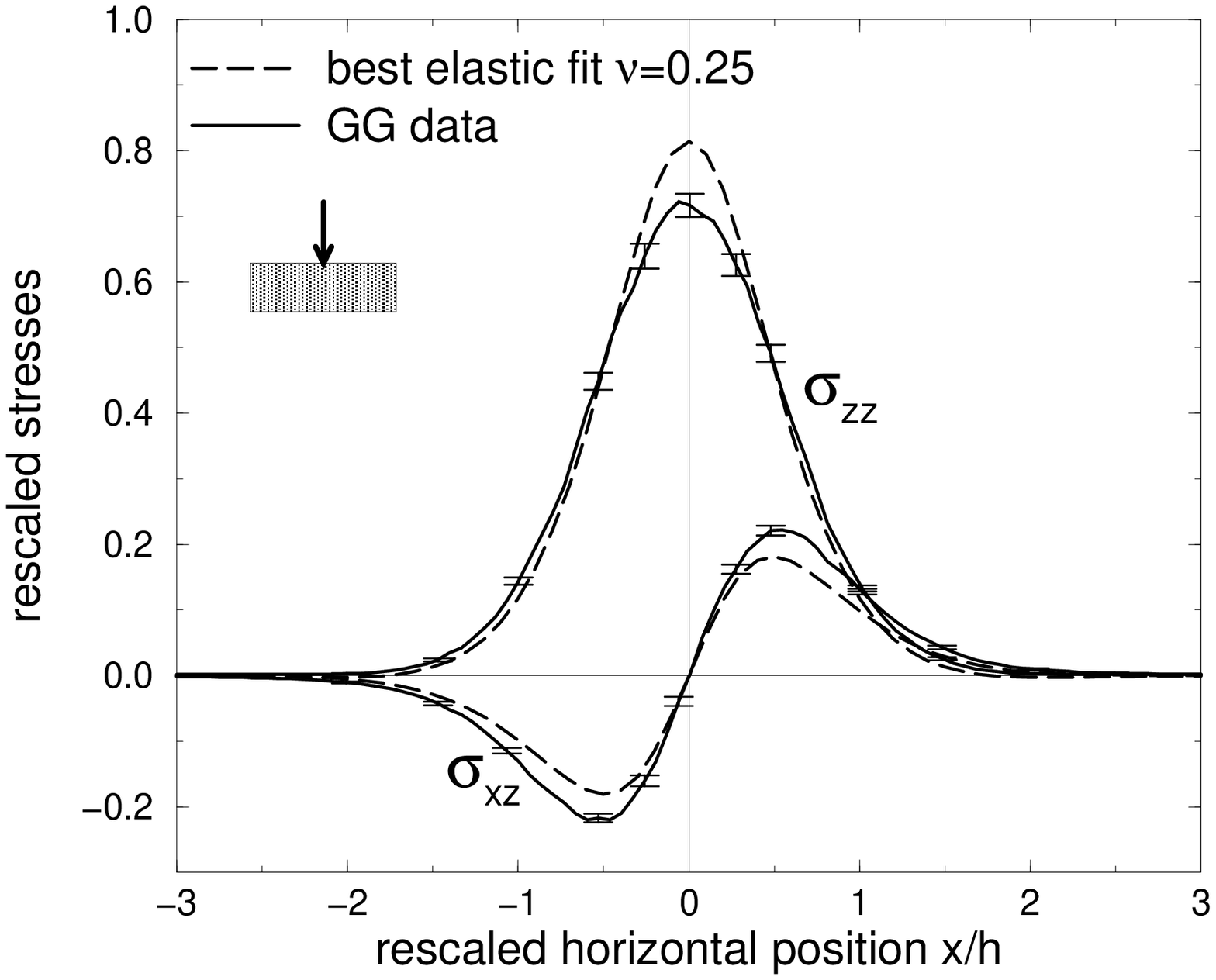},width=0.48\linewidth,angle=0}
\hfill
\epsfig{file={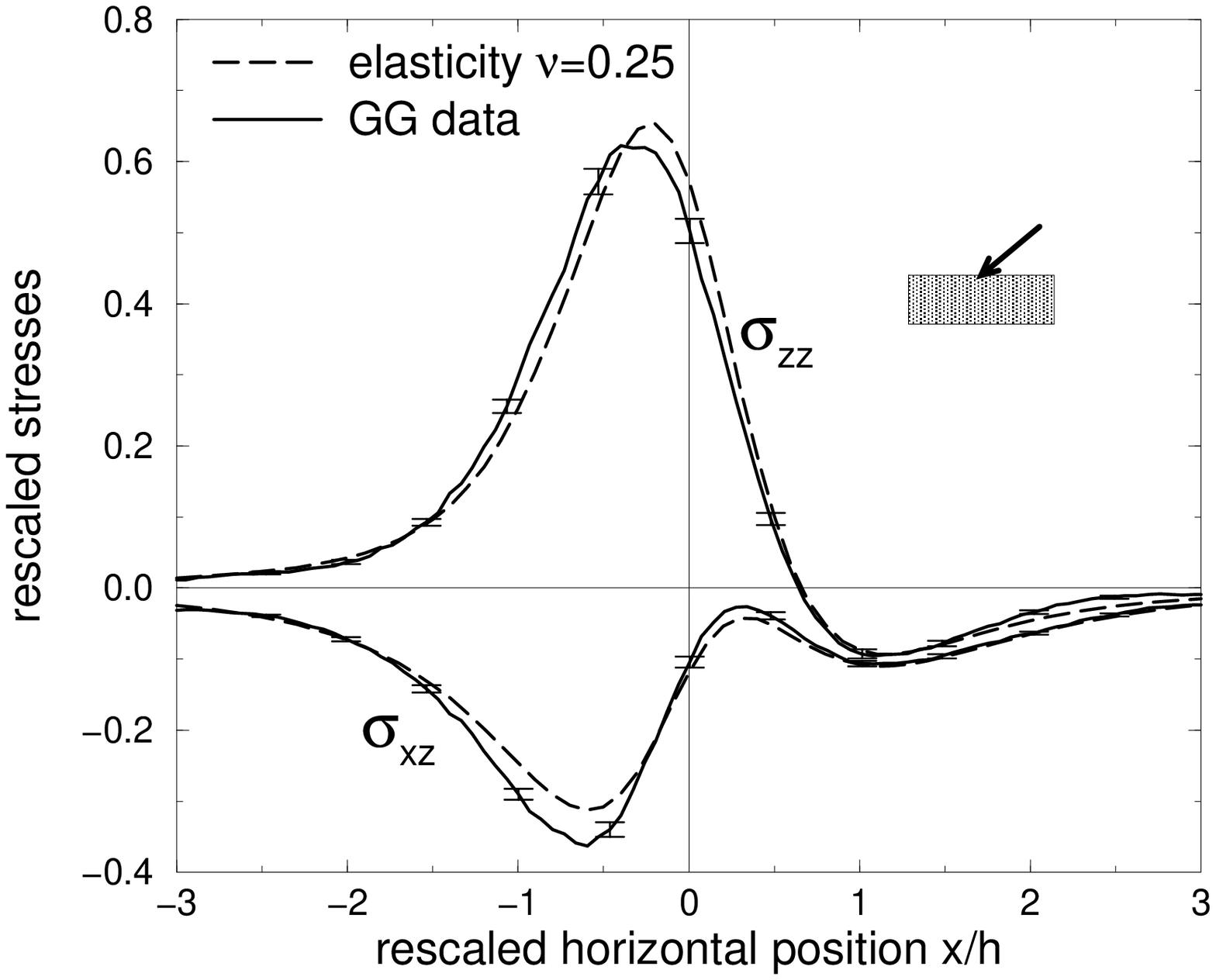},width=0.48\linewidth,angle=0}
\end{center}
\caption{Equivalent of figure \protect\ref{fig:fitsijRL} for the case
of the grain-by-grain preparation. Here, elastic predictions of
isotropic elasticity for both vertical and non-normal extra forces are
much less good.}
\label{fig:fitsijGG}
\end{figure}

In the same vein as the experiments discussed above, \cite{SRCCL01},
we performed fits of the numerical profiles by minimizing the
quadratic distance between the data and the theoretical predictions,
including the weight of the uncertainties. Recall that the fitting
parameter is the Poisson coefficient $\nu$. 
As shown in figure \ref{fig:fitsijRL}, the RL profiles can be well reproduced
with $\nu=0.28(2)$ (best fit) when the extra force $F_0$ is vertical. This
value of the Poisson ratio also fits well the stresses in response to an
overload inclined at $45^\circ$. In the case of the GG preparation (figure
\ref{fig:fitsijGG}), the best fit of the stress responses to a vertical $F_0$
($\nu=0.25(2)$) is not as well adjusted to the numerical data. The discrepancy
between isotropic elasticity and the GG data is also clearly visible for the
inclined $F_0$ situation. Using a range of directions for the applied extra
force provides a good test to identify a medium showing weak anisotropy such
as occurs for preparations under gravity.

\section{Discussion and conclusion}

Here we presented a summary of recent results based on the measurement
of stress responses to localized forces. These results are basically
consistent with an elastic description. We have shown experimentally
for 2D model granular assemblies and for 3D sand piles that the
response function is a useful probe of constitutive properties of the
granular material. Indeed, this type of probe is a non intrusive
method that could replace or supplement standard acoustic techniques,
since it is well adapted to weakly cohesive material under low
confining pressures where sound propagation methods are more difficult
to implement.

We have shown that the response function is sensitive to small
differences in the preparation history and also that its shape may
change significantly after a quasi-static deformation. Using numerical
simulations, we associate these macroscopic differences to different
microscopic features which appear in the angular distribution of
contacts. Nevertheless, exactly how these features would
quantitatively emerge at the macroscopic level and how they are likely
to change due to an external loading path is still a challenging and
unanswered question.

Finally, we propose a new path of systematic measurements where the
normal and shear stress profiles are obtained for various directions
of the applied localized force. A collection of such data would yield
crucial information on the structure evolution for a given loading history
and, therefore, provide a severe test of any model such as anisotropic
elasticity models.

\noindent
\rule[0.1cm]{3cm}{1pt}

We thank C.~Goldenberg, I.~Goldhirsch, J.~Jenkins, S.~Luding and
J.~Snoeijer for fruitful discussions.
A.P.F. Atman's present address is Departamento de F\'\i sica, Instituto de
Ci\^encias Exatas, Universidade Federal de Minas Gerais, C.P. 702, 30123-970,
Belo Horizonte, MG, Brazil.
P. Brunet's present address is Royal Institute of Technology -- Department of
Mechanics, 10044 Stockholm, Sweden.
The LPMMH is UMR 7636 of the CNRS.
The work of RPB was supported by NSF grants DMR-0137119 and DMS-0204677.

\vspace*{1cm}


\end{document}